\documentstyle[12pt,prl,aps,psfig]{revtex}

\begin{document}
\draft
\title{Reduction of Magnetic Noise \\in Magnetic Resonance Force Microscopy}
\author{G.P. Berman$^1$, V.N. Gorshkov$^{1,2}$, and V.I. Tsifrinovich$^3$}
\address{$^1$ Theoretical Division, Los Alamos National
Laboratory, 
Los Alamos, New Mexico 87545}
\address{$^2$ Department of Physics, Clarkson University, Potsdam, NY 13699}
\address{$^3$ IDS Department, Polytechnic University, Brooklyn, NY 11201}
\maketitle
\vspace{5mm}
\begin{abstract}
We study the opportunity to reduce a magnetic noise produced by a uniform cantilever with a ferromagnetic particle in magnetic resonance force microscopy (MRFM) applications. We demonstrate theoretically a significant reduction of magnetic noise and the corresponding increase of the MRFM relaxation time using a nonuniform cantilever.
\end{abstract}
\section{Introduction}
A micro-mechanical cantilever has found enormous application in atomic scale microscopy. In particular, a cantilever with a ferromagnetic particle became the main tool in magnetic resonance force microscopy (MRFM) \cite{1}. One of the important problems in MRFM is the magnetic noise on the Rabi frequency caused by the thermal excitations of the high-frequency cantilever modes \cite{2,3,4}. 
In this paper, we study the suppression of the high-frequency magnetic noise using a nonuniform cantilever. 

There are, at least, three ways to reduce the magnetic noise using a nonuniform cantilever \cite{3,5,6}:\\
1) The increase of the cantilever mass $m_c$ causes the decrease of the amplitude of the thermal cantilever vibrations $A_n$ as $m_c(A_n\omega_n)^2/2=k_BT$ for a mode with the frequency $\omega_n$, where $T$ is the temperature.\\
2) One can try to reduce the value of the cantilever eigenfunctions $f_n(x)$ near the location of the ferromagnetic particle.\\
3) One can increase the gap $\Delta\omega_n$ between the eigenfrequencies of the cantilever in the region of the Rabi frequency.

Below we study the effect of the inhomogeneity on the eigenfunctions $f_n(x)$, frequency gap $\Delta\omega_n$, and the resulting change of the MRFM relaxation time. We show that the magnetic noise can be significantly reduced using a nonuniform cantilever. 
\section{The model of nonuniform cantilever}
To describe the cantilever vibrations, we use the well-known equation of motion for the cantilever \cite{7}:
$$
\rho S(x){{\partial^2z}\over{\partial t^2}}=-E{{\partial^2}\over{\partial x^2}}\Bigg(I(x){{\partial^2z}\over{\partial x^2}}\Bigg).\eqno(1)
$$
Here $\rho$, $S(x)$, $E$ are the density, the cross-section area, and the Yonng's modulus of the cantilever, $I(x)=w_ct^3_c/12$, $w_c$ and $t_c$ are the width and the thickness of the cantilever. The boundary conditions are 
$$
z(0)=\Bigg({{\partial z}\over{\partial x}}\Bigg)_0=\Bigg({{\partial^2 z}\over{\partial x^2}}\Bigg)_{l_c}=\Bigg({{\partial^3 z}\over{\partial x^3}}\Bigg)_{l_c}=0,\eqno(2)
$$
where $l_c$ is the length of the cantilever ($t_c\ll w_c\ll l_c$).

The eigenfunctions $f_n(x)$ and the eigenfrequencies $\omega_n$ of the cantilever satisfy the equation
$$
\rho S(x)f_n(x)\omega^2_n=E{{\partial}\over{\partial x^2}}\Bigg(I(x){{\partial^2}\over{\partial x^2}}f_n(x)\Bigg).\eqno(3)
$$
The conditions of orthogonality can be easily derived from (3) 
$$
\int^{l_c}_0S(x)f_n(x)f_m(x)dx=0.\eqno(4)
$$
We normalize the eigenfunctions in a standard way
$$
{{\int^{l_c}_0S(x)f^2_n(x)dx}\over{\int^{l_c}_0S(x)dx}}=1.\eqno(5)
$$
To find the eigenfunctions $f_n(x)$ of the nonuniform cantilever we use the expansion over the eigenfunctions $\varphi_n(x)$ of the uniform cantilever
$$
f_n(x)=\sum_{k=1}^Mc_{kn}\varphi_k(x).\eqno(6)
$$
The eigenfunctions $\varphi_k(x)$ satisfy the same boundary conditions as $f_n(x)$ and the normalization condition
$$
{{1}\over{l_c}}\int^{l_c}_0\varphi^2_n(x)dx=1.\eqno(7)
$$
We take the number of basis functions $M=50$ which provides the accurate approximation for the eigenfunctions of the nonuniform cantilever. In particular, we have verified the conditions of orthogonality (4) for the normalized eigenfunctions $f_n(x)$: the value of the integral in the left side of Eq. (4) did not exceed 0.02. In our computations for a given value of $n$ we have found the eigenvectors $c_{kn}$ $(1\le k\le M)$ of the matrix $\alpha_{km}$
$$
l_c\omega^2_nc_{kn}=\sum_{m=1}^M\alpha_{km}c_{mn},\eqno(8)
$$
$$
\alpha_{km}={{E}\over{\rho}}\int^{l_c}_0\Bigg[{{\partial^2}\over{\partial x^2}}\Bigg({{\varphi_k(x)}\over{S(x)}}\Bigg)\Bigg]I(x){{\partial^2\varphi_m(x)}\over{\partial x^2}}dx.
$$
\section{Numerical simulations}
We have considered the nonuniform increase of the thickness of the cantilever
$$
t^\prime_c(x)=t_c\{1+\gamma\exp[-(x-x_0)^2/\delta^2]\},\eqno(9)
$$
where $t_c$ is the thickness of the ``unperturbed'' cantilever, $(1+\gamma)t_c$ is the maximum thickness which is achieved at the point $x=x_0$, and $\delta$ is the characteristic size of inhomogeneity. For $t_c\ll w_c\ll l_c$ the eigenfrequencies and the eigenfunctions depend on the ratio $l_c/t_c$ which was chosen approximately 33 in our simulations. In some cases, for $x>x_0$ we take the constant value of the thickness: $t^\prime_c(x)=t_c(1+\gamma)$. (See captions to Figs 1,4,5.)

Typically, a ferromagnetic particle in MRFM experiments is placed near the cantilever tip. Our simulations show that in this case one should increase the thickness of the cantilever near its tip in order to provide the maximum possible reduction of the high-frequency eigenfunction value near the tip. 

\begin{figure}[t]
\centerline{\psfig{file=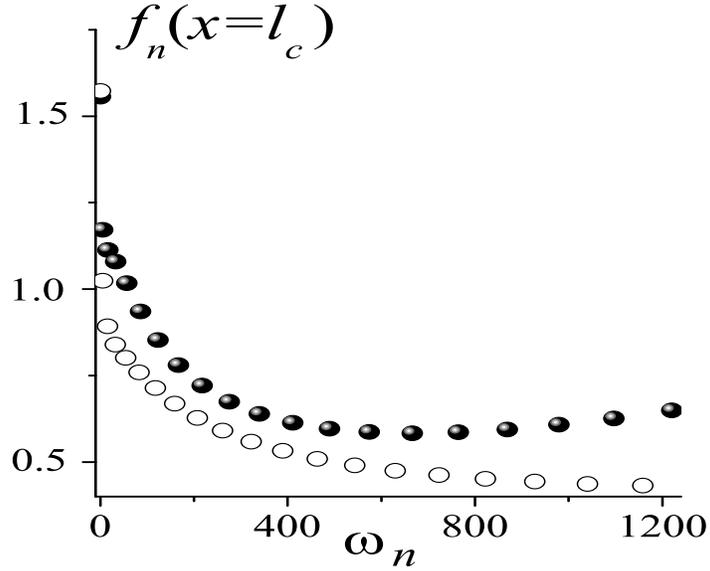,width=11cm,height=9cm,clip=}}
\vspace{4mm}
\caption{Dependence of the values of eigenfunctions $f_n(l_c)$ on the eigenfrequency $\omega_n$. Dots correspond to $\gamma=4$, $x_0=0.95$, $\delta=0.05$. The mass of the cantilever (in terms of the mass of the ``unperturbed'' uniform cantilever) $m_c=1.38$. The fundamental frequency $\omega_1=0.65$. Circles correspond to $\gamma=6$, $x_0=0.99$, $\delta=0.05$, $m_c=1.325$, $\omega_1=0.68$. For $x>x_0$ the cantilever thickness is constant.}
\label{fig:1}
\end{figure}

Fig. 1 shows the values of the eigenfunctions $f_n(x)$ near the tip $(x=l_c)$ as a function of the eigenfrequency $\omega_n$. In Fig. 1 and below the eigenfrequencies of the modes $\omega_n$ are given in the units of the fundamental frequency $\omega_1$ of the ``unperturbed'' uniform cantilever. The values $x_0$ and $\delta$ are given in units $l_c$. For the uniform cantilever the values of all eigenfunctions near the cantilever tip are the same: $|\varphi(l_c)|=2$ \cite{7}. 

We have found that even a more significant effect in the noise reduction can be achieved if the ferromagnetic particle is located at some distance from the cantilever tip. In this case, the increase of the cantilever thickness should be centered at a position of the ferromagnetic particle. 
Fig. 2 demonstrates the reduction of the eigenfunctions $f_{11}(x)$ and $f_{12}(x)$ in the region of the inhomogeneity. Circles in Fig. 3b show the values of the eigenfunctions at the center of the inhomogeneity $x=x_0<l_c$. 
One can see that the values of $f_n(x_0)$ may become very small. At the same time, the values $f_n(l_c)$ increase in comparison to the uniform cantilever (squares in Fig. 3). 

\begin{figure}[t]
\centerline{\psfig{file=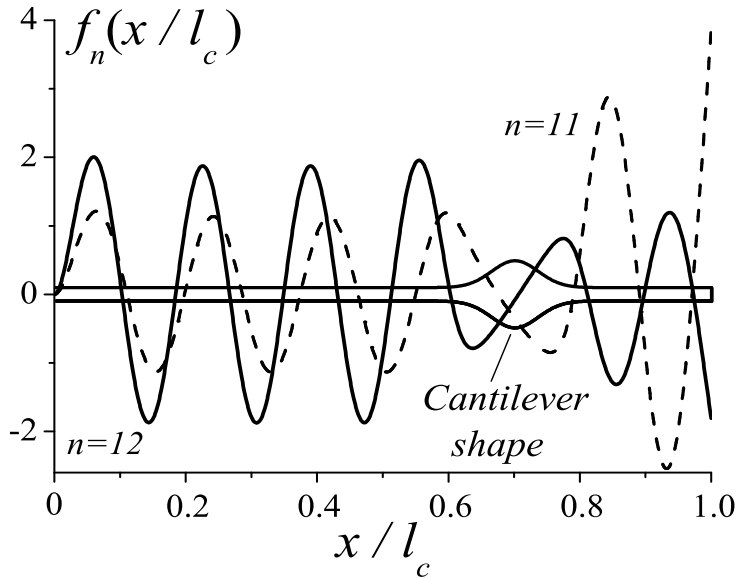,width=11cm,height=9cm,clip=}}
\vspace{4mm}
\caption{ Eigenfunctions $f_n(x/l_c)$ for $n=11$ and $n=12$ (the corresponding eigenfrequencies $\omega_{11}=362$ and $\omega_{12}=416$). The values of parameters: $\gamma=4$, $x_0=0.7$, $\delta=0.05$.}
\label{fig:2}
\end{figure}

Fig. 4 demonstrates the same features as Fig. 3 for the cantilever of a different shape. One can see that in the frequency region $\omega_{eff}$ the values $f_n(x_0)$ are very small: $f_n(x_0)\ll 1$. 
For a ``heavily loaded'' cantilever with the mass increase greater than 100\%, the reduction of values of the eigenfunctions is not significant, however, the increase of the gap between the eigenfrequencies is rather large \cite{6}. 
Fig. 5 demonstrates both the change of the values of eigenfunctions near the tip and the ``repulsion'' of the eigenfrequencies, for a heavily loaded cantilever. 
Using the method described in \cite{3} we computed the relaxation time for the MRFM signal $\Delta T/\Delta T_0$, where $\Delta T$ is the shift of the cantilever period caused by the spins of the sample, and $\Delta T_0$ is the initial value of $\Delta T$. (As in \cite{3}, we assume that the relaxation is caused by the thermal vibrations of the high-frequency cantilever modes.) 

\begin{figure}[t]
\centerline{\psfig{file=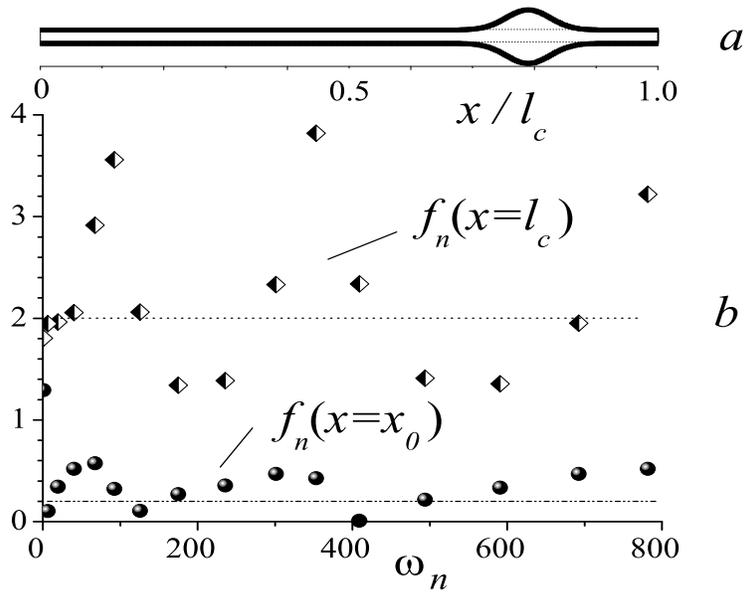,width=11cm,height=9cm,clip=}}
\vspace{4mm}
\caption{(a) The cantilever shape; (b) the values of the eigenfunctions at the center of the inhomogeneity $x=x_0$ (circles) and near the cantilever tip $x=l_c$ (squares). The values of parameters: $\gamma=3$, $x_0=0.79$, $\delta=0.05$.}
\label{fig:3}
\end{figure}

\begin{figure}[t]
\centerline{\psfig{file=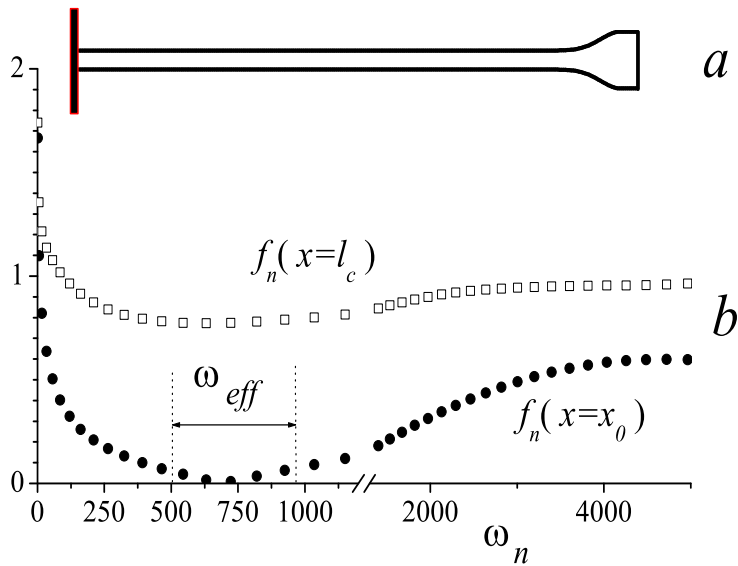,width=11cm,height=9cm,clip=}}
\vspace{4mm}
\caption{The same as in Fig. 3 but for $\gamma=2$, $x_0=0.97$, $\delta=0.05$.
For $x>x_0$ the cantilever thickness is constant.}
\label{fig:4}
\end{figure}

Fig. 6 shows the typical decay of the MRFM signal for the following parameters. 
The amplitude of the cantilever tip vibrations is 15 nm, the effective spring constant and the fundamental frequency of the ``unperturbed'' uniform cantilever are 0.014 N/m and 21 kHz, correspondingly, the Rabi frequency is 8.2 MHz, the magnetic moment of the ferromagnetic particle is $1.5\times 10^{-12}$ J/T, the radius of the ferromagnetic particle is 700 nm, the distance from the bottom of the ferromagnetic particle to the center of the resonant slice is 875 nm, the cantilever quality factor is $2\times 10^4$. The cantilever oscillates perpendicular to the surface of the sample. To reduce the computational time, we took the room temperature.

\begin{figure}[t]
\centerline{\psfig{file=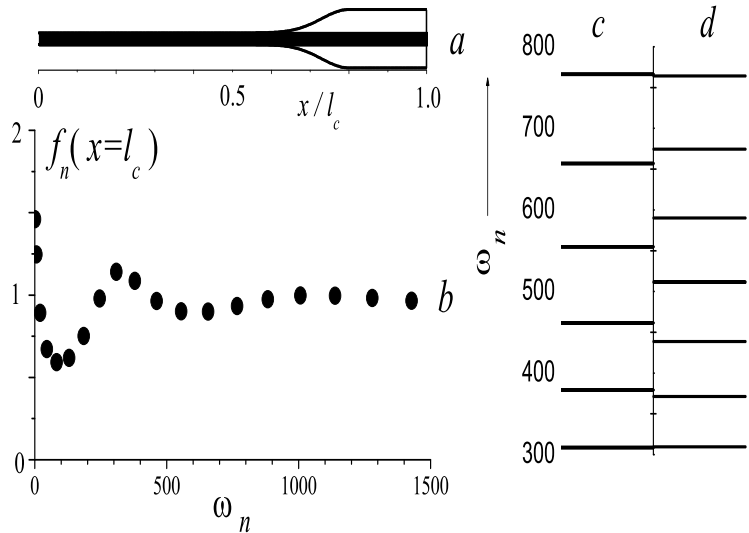,width=11cm,height=9cm,clip=}}
\vspace{4mm}
\caption{The values of eigenfunctions near the cantilever tip and the eigenfrequencies for a heavily loaded cantilever; (a) the cantilever shape; (b) the values of eigenfunctions near the tip; (c) the eigenfrequencies of the cantilever; (d) the eigenfrequencies of the ``unperturbed'' uniform cantilever. The values of parameters: $\gamma=4$, $x_0=0.8$, $\delta=0.1$, $m_c=2.15$, $\omega_1=0.5$. For $x>x_0$ the cantilever thickness is constant.}
\label{fig:5}
\end{figure}

Curve (a) in Fig. 6 corresponds to the uniform cantilever with the ferromagnetic particle near the cantilever tip, curve (b) corresponds to the cantilever with the ferromagnetic particle and the inhomogeneity near the tip ver tip (dots in Fig. 1 show the eigenfunctions for this cantilever), curves (c) and (d) correspond to the cantilever with the ferromagnetic particle  placed at some distance from the tip: $x=x_0< l_c$ (curve (c) is for the cantilever shown in Fig. 3a, and curve (d) is for the cantilever shown in Fig. 4a). The ratio of the relaxation times for these four cases is: 1:10:63:140. 

For a heavily loaded cantilever shown in Fig. 5 with the ferromagnetic particle placed near the cantilever tip ($x=l_c$) the decay of the MRFM signal is almost the same as that described by the curve (d) in Fig. 6. 

\begin{figure}[t]
\centerline{\psfig{file=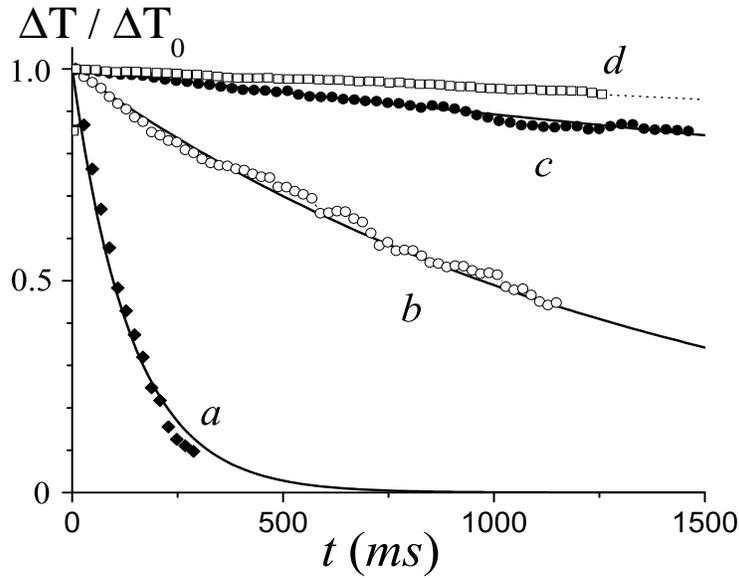,width=11cm,height=9cm,clip=}}
\vspace{4mm}
\caption{Typical decay of the MRFM signal for four cases: (a) the uniform cantilever with the ferromagnetic particle near the tip; (b) the nonuniform cantilever with the ferromagnetic particle near the tip ($\gamma=4$, $x_0=0.95$, $\delta=0.05$); (c) the nonuniform cantilever with the ferromagnetic particle at $x=x_0<l_c$ ($\gamma=3$, $x_0=0.79$, $\delta=0.05)$; (d) the same for $\gamma=2$, $x_0=0.97$, $\delta=0.05$.}
\label{fig:6}
\end{figure}

\section*{Conclusion}

We studied the opportunity to reduce the magnetic noise using a nonuniform cantilever in MRFM applications. Our results can be formulated as follows:\\
1) Small inhomogeneity of the cantilever thickness (with the mass change less than 50\%) near the cantilever tip, can provide a ten-hold increase 
 of the MRFM relaxation time.\\
2) Greater suppression of the magnetic noise can be achieved if the ferromagnetic particle is located at some distance from the cantilever tip. The same results can be achieved using a heavily loaded cantilever (with the mass change more than 100\%).  
\section*{Acknowledgments}
This work  was supported by the Department of Energy under the contract W-7405-ENG-36 and DOE Office of Basic Energy Sciences, by the National Security Agency (NSA), by the Advanced Research and Development Activity (ARDA), and by the DARPA Program MOSAIC. 
{}

\begin{thebibliography}{}
%
\bibitem{1}
J.A. Sidles, Rev. Mod. Phys., {\bf 67}, 249 (1995).
%
\bibitem{2}
B.C. Stipe, H.J. Mamin, C.S. Yannoni, T.D. Stowe, T.W. Kenny, D. Rugar, 
Phys. Rev. Lett., {\bf 8727}, 7602 (2001).
%
\bibitem{3}
G.P. Berman, V.N. Gorshkov, D. Rugar, V.I. Tsifrinovich, quant-ph/0303171,  accepted by Phys. Rev. B, (2003).
%
\bibitem{4}
D. Mozyrsky, I. Martin, D. Pelekhov, P.C. Hammel, Appl. Phys. Lett., {\bf 82},
1278 (2003).
%
\bibitem{5}
G.P. Berman, talk on the DARPA MOSAIC Review, October, 16-18, 2002, Los Angeles, CA.
%
%
\bibitem{6}
B.W. Chui, Y. Hishinuma, R. Budakian, H.J. Mamin, T.W. Kenny, D. Rugar, in publication.
%
\bibitem{7}
L.D. Landau and E.M. Lifshitz, {\it Theory of Elasticity},  Oxford,  New York, Pergamon Press, 1986.
1986.
%
\end{thebibliography}
\end{document}